\documentclass[showpacs,prl,aps,superscriptaddress,twocolumn,floatfix]{revtex4}
\usepackage{graphicx,float,amsmath,amssymb}

\begin{document}

\title{Dipole Oscillations of a Bose-Einstein Condensate in Presence of
  Defects and Disorder}

\author{M. Albert}
\author{T. Paul}
\author{N. Pavloff}
\author{P. Leboeuf}
\affiliation{Laboratoire de Physique Th\'eorique
et Mod\`eles Statistiques, CNRS, 
Universit\'e Paris Sud, UMR8626, 91405 Orsay Cedex, France}
\begin{abstract}
We consider dipole oscillations of a trapped dilute Bose-Einstein condensate
in the presence of a scattering potential consisting either in a localized
defect or in an extended disordered potential. In both cases the breaking of
superfluidity and the damping of the oscillations are shown to be related to
the appearance of a nonlinear dissipative flow. At supersonic velocities the
flow becomes asymptotically dissipationless.
\end{abstract}


\maketitle


One of the most spectacular consequences of phase coherence and interactions
in condensed matter is superfluidity, a direct manifestation of which is
the capacity of a fluid to move without dissipation. 
According to the standard Landau
criterion, the superfluidity (SF) of a uniform flow of, e.g., liquid He$^4$, or 
a Bose Einstein condensate (BEC) will be broken if an obstacle moves through 
the fluid with a speed higher than a critical velocity 
$v_c^{\rm L}=\mbox{min}\,\{E(p)/p\}$ where $E(p)$ is the dispersion
relation of elementary excitations of momentum $p$.
Though this property has been explicitly checked in He$^4$ \cite{All76} and in
a BEC flow \cite{Chik00}
in presence of small impurities, experiments in superfluid $^4$He (see, e.g.,
\cite{Var91}) and more recently in BEC \cite{Raman99,Engels07} have shown that
the critical velocity for breaking SF is generically lower than $v_c^{\rm L}$,
due to phase slips induced by vortex (or soliton) emission,
as originally proposed by Feynman \cite{Fey55}.  

Collective oscillations of BEC confined by harmonic traps offer new
opportunities to explore the central question of the breaking of SF and of the
origin of drag and dissipation in quantum liquids and gases. In a recent
series of experiments, damping of the oscillations (such as dipole or
quadrupole) in the presence of a single localized scatterer 
\cite{FortPRL05}, and disordered
\cite{Lye05,Chen07} or quasiperiodic \cite{LyeRapid07} superimposed
potentials has been used to investigate different dynamical regimes,
including the possibility of a Bose glass, (Anderson) localization or other
possible phases. These investigations have clearly shown the experimental
relevance of analyzing transport properties of BEC {\it via} the damping of
collective excitations. However the connexion of the damping with localization
properties still remains to be clarified.

Our purpose here is to provide a global analysis of the phase diagram related
to the damping of dipole oscillations in the presence of a single localized
scatterer or a disordered potential. We consider the regime where the
experiments have been realized up to now, i.e., a quasi-1D geometry
where the chemical potential is larger than the typical amplitude of the
perturbing potential.  The reason why we treat together the localized defect
and the random potential is that, qualitatively, several of the main features
of the dynamics are contained in the former case. Its analysis therefore
facilitates the comprehension of the latter, and stresses the generic aspects,
leading to a unified picture of dissipation. We find that in both cases there
exist SF undamped oscillations at small amplitudes. As the amplitude of
oscillation (or the typical size of the perturbation) increases, the system
enters a dissipative regime where solitons and phonon-like excitations are
emitted (a regime recently studied experimentally in Ref.\cite{Engels07} for a
moving localized obstacle). Though the system preserves its phase coherence,
the collectivity of the center of mass motion, and therefore the amplitude of
the dipole oscillations, diminishes \cite{Lye05,Chen07}. In the case of a
superimposed disordered potential, this dissipative or resistive phase, where
non-linearities of the system play a crucial role, has no relation with
(Anderson) localization.

\begin{figure}
\includegraphics*[width=0.9\linewidth,angle=0]{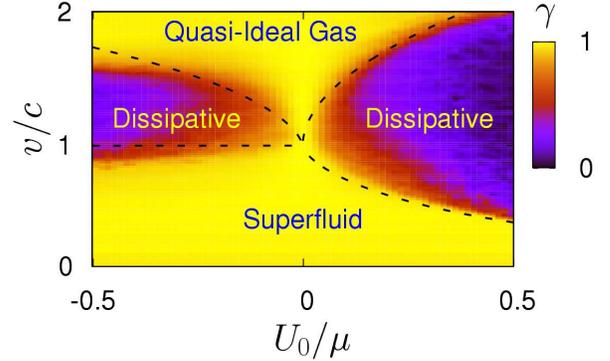}
\caption{(color online) Dynamical regimes for dipole oscillations in
  presence of a localized Gaussian defect. The plot represents the
fluidity factor $\gamma$ (see the text) computed after a time
  $t_f=25\times 2 \pi/\omega_x$. The yellow (light gray) region correspond to
zero damping ($\gamma=1$).
The dashed lines are analytic determinations of the frontiers between the
different regimes (see text).} \label{F1}
\end{figure}

The system considered is a weakly interacting BEC confined in a cylindrically
symmetric 3D harmonic potential
$m(\omega^2_{\perp}r^2_{\perp}+\omega^2_x x^2)/2$ in presence of
an additional potential $U(x)$. In the limit of a highly anisotropic trap,
$\omega_{\perp}\gg \omega_x$, the transverse confinement is such that the
quasi-$1$D regime can be reached. 
It is important to note that for moderate $U(x)$ (even a disordered one)
the phase coherence of the system is preserved
as demonstrated in Refs. \cite{Chen07,Cle08}.
The system is thus accurately
described by a 1D order parameter $\psi(x,t)$, depending on a single spatial
coordinate $x$ along the axial direction of the trap. $\psi(x,t)$ obeys the
nonlinear Schr\"odinger equation \cite{Jack98,Leb01}
\begin{equation} \label{10}
i \hbar \frac{\partial \psi}{\partial t}=
\left[-\frac{\hbar^2}{2m}\frac{\partial^2}{\partial x^2} + 
\frac{m}{2}\omega^2_x\,x^2+ U(x)+2\hbar\omega_\perp (a n)^{\nu} \right]\psi \; .
\end{equation}
Here, $n(x,t)\equiv |\psi(x,t)|^2$ is the condensate density per unit of
longitudinal length and 
$a>0$ is the $3$D s-wave scattering length.  In the low density regime (LDR,
$a n\ll 1$) the density profile in the transverse direction is Gaussian-shaped
and $\nu=1$, whereas $\nu=1/2$ in the opposite high density regime (HDR, $a
n\gg 1$) where the Thomas-Fermi approximation holds for the transverse degree
of freedom. Note that Eq. (\ref{10}) does not account for transverse
excitations which may be relevant in the HDR.
We checked that it nonetheless gives an excellent
account of the experimental result on dipole oscillations
of the Florence and Rice groups
\cite{Lye05,Chen07} performed in the HDR.

After preparing the condensate in the ground state of the harmonic trap (with
density $n_0(x)$, chemical potential $\mu$ and, at the center of the cloud,
speed of sound $c$), dipole oscillations were excited by a sudden displacement
$d_0$ of the harmonic potential. For $U(x)\equiv 0$ the center of mass
oscillates freely with frequency $\omega_x$, and acquires a velocity
$v=\omega_x d_0$ when passing through the origin. The time evolution of the
density reads $n (x,t)=n_0(x-X_t)$ where $X_t= d_0\cos(\omega_x t)$ is the
position of the center of mass. For a finite $U(x)$, which is turned on
simultaneously with the sudden displacement of the trap,
$X_t=\frac{1}{N}\int_\mathbb{R} dx\,x\,n(x,t)$ is computed numerically up to a
time $t_f$ chosen such that $X_{t>t_f}$ assumes an oscillatory pattern of
roughly constant amplitude which we denote $d_f$.  In order to measure the
damping of the dipole oscillations we define a fluidity factor $\gamma=d_f /
d_{0}$ ($\gamma=1$ in the absence of damping and $\gamma\to 0$ for strong
damping).

{\sl Localized defect.} We start by considering a Gaussian-shaped defect
$U(x)=U_0 \exp(-x^2/2\sigma^2)$.
The fluidity factor $\gamma$ is plotted in Fig.~\ref{F1} as a function of the
normalized defect strength $U_0/\mu$ and velocity $v/c=\omega_x d_0/c$.
The numerical calculations were performed for a BEC in the
LDR with chemical potential $\mu=40\, \hbar \omega_x$ and aspect ratio
$\omega_x/\omega_{\perp}$ selected to have a Thomas-Fermi like density-profile
along the axial direction. In this case
$n_0(x)={\mathcal N}_{\nu} \, \Theta(L-|x|) \, [L^2-x^2]^{1/\nu}$ where
the factor ${\mathcal N}_{\nu}$ normalizes the density to the number of atoms
$N$, $\Theta (x)$ is the Heavyside step function
and $L=\sqrt{2\mu/m\omega_x^2} $ is half the longitudinal size of the
condensate. 
The parameters are $N=1.5\times 10^4$ $^{87}$Rb
atoms, $\omega_x=2\pi\times 9$ s$^{-1}=\omega_{\perp}/10$
and $\sigma=\xi= 0.28$ $\mu$m,
where $\xi=\hbar/\sqrt{2 m \mu}$ is the healing length at the center of the
condensate. The qualitative structure of Fig.~\ref{F1} is generic and does
not depend on the specific values of $\sigma$ and $ \mu $, and
is also observed in the HDR ($\nu=1/2$).

\begin{figure}
\includegraphics*[width=0.95\linewidth,angle=0]{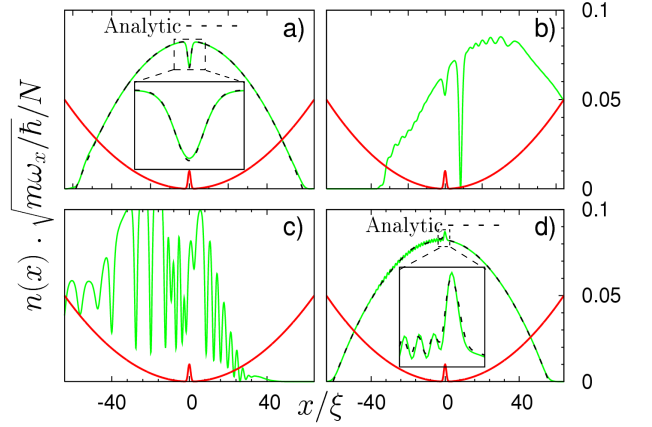}
\caption{(color online) Density profile after a time $t=3/4 \times 2
  \pi/\omega_x$ for $U_0/\mu=0.24$ at different initial velocities: a)
  $v/c=0.1$, b) $v/c=0.67$, c) $v/c=1.2$, d) $v/c=2.5$. The confining
  potential is represented as a full (red) curve.  Insets blow up the
  density around the defect. (Other parameters as in Fig.~\ref{F1})}
\label{F2}
\end{figure}

In the deep subsonic limit $v/c \ll 1$, the Gaussian scatterer induces no
observable damping of the dipole oscillations. Numerical results show that
the oscillating condensate is only locally perturbed in the vicinity of the
defect: a dip or a peak appear in the condensate density for $U_0 > 0$ and
$U_0 < 0$, respectively (see Fig.~\ref{F2}(a)). These are characteristic
features of a superfluid flow, with no energy dissipation, nor drag exerted
\cite{Pav02}, and with no damping of the oscillations (perfect transmission
through the scatterer potential).  In this regime, a perturbative treatment of
equation (\ref{10}) and a local density approximation yield a condensate
density of the form
\begin{equation}\label{eq}
n(x,t) = n_0 (x-X_t) \left[1+\delta n(x,t) \right]\; ,
\end{equation} 
with
\begin{equation} \label{20}
\delta n(x,t)=-\frac{2m}{\hbar^2\kappa}
\int_{-\infty}^{+\infty}dy\,e^{-\kappa|x-y|}\,U(y) \; ,
\end{equation}
where $\kappa=\frac{2 m}{\hbar} |c_0^2-\dot{X}_t^2|^{1/2}$, $c_0$ being
the unperturbed local sound velocity: $m\,c_0^2 = 2\nu\hbar\omega_\perp
[a\,n_0(x-X_t)]^\nu$.  The accuracy of this approximation is shown in
Fig.~\ref{F2}(a); it is well justified if $m U_0 \sigma/(\hbar^2 \kappa) \ll
1$ when $\kappa\sigma\ll 1$, or $m U_0 /(\hbar \kappa)^2 \ll 1$ when
$\kappa\sigma\gg 1$ \cite{Leb01}, and if $\sigma\ll L$.

For weak defect potentials, $|U_0| \ll \mu$, and if the TF size $L$ is large
compared to the dipole oscillation amplitude, the center of mass position
$X_t$ can be computed analytically. To lowest order, the solution of the
small-amplitude linearization yields $X_t= d_0 \cos[(\omega_x+\delta\omega)t]$
where the defect-induced frequency shift reads
\begin{equation} \label{50}
\delta \omega = \frac{-1}{2 m \omega_x}\int_{-\infty}^{+\infty}dx\, 
\frac{d n_0(x)}{dx}\, \frac{d U(x)}{dx} \; .
\end{equation}
This gives $\delta \omega=- (4-3\nu)\frac{3\,U_0\, \sigma}{8\,\mu^{3/2}} \,
\omega_x^2\, \sqrt{\pi m}$ for a Gaussian defect, in excellent agreement with
our numerical results. Hence the analytical evaluations of the density profile
(\ref{eq}) and of the center of mass motion confirm the superfluid behavior of
the oscillations in the deep subsonic regime.

The situation changes as the velocity increases at fixed $U_0/\mu$ or as
$U_0/\mu$ is increased at constant velocity. In the former case, at some
critical velocity $v_c$ ($\leq c$), that depends on the strength of the defect
potential, the system looses SF, damping is observed and the fluidity factor
diminishes. In the spirit of Landau's criterion 
we identify the border between the SF and this ``dissipative region''
as the locus of points where the maximum local condensate velocity $v(x,t)$
equals the local speed of sound $c(x,t) = (2\nu \hbar\omega_\perp/m)^{1/2}
[a\,n(x,t)]^{\nu/2}$ \cite{Hakim97}. The former can be computed from mass
conservation and Eq.~(\ref{eq}). For an impurity localized at $x=0$
this yields
\begin{equation} \label{70}
\frac{v_c}{c}=[1+\delta n_c]^{1/2+1/\nu} \; \text{if}\;\; U_0>0 \; ,\;
 \frac{v_c}{c}=1 \; \text{if}\;\; U_0<0.
\end{equation}
where $\delta n_c$ is the factor 
$\delta n(x,t)$ [Eq. (\ref{20})] evaluated when $x=X_t=0$.
Fig.~\ref{F1} shows that these estimates coincide well with the numerical
findings.

When the system enters the dissipative regime, one or a few gray
solitons detach from the defect during the first oscillations, as well as some
phonon-like excitations (see Fig.~\ref{F2}(b)).
As time goes on, the interactions of the solitons among them, with the
defect and with the phonon-like excitations produce time-dependent
fluctuations of the shape. During this process the condensate does not loose
phase coherence, but the center of mass motion looses collectivity, part of
the kinetic collective energy being transformed into density fluctuations. The
damping process continues until the center of mass velocity becomes comparable
with the critical velocity. Thereupon, though presenting local density
fluctuations, the amplitude of the oscillations remain constant in time.
Deeper in the dissipative regime, an increased emission of gray solitons and
phonon-like excitations is observed, leading to a massive distortion of the
initial condensate profile (see Fig. \ref{F2}(c)).  The BEC enters a strongly
irregular time-dependent regime, the collectivity of the
dipole motion is totally lost, and the damping 
increases drastically.

Finally, at sufficiently high supersonic velocities, a different phase is
reached where the damping tends again to zero ($\gamma\to 1$).
In this regime, that we denote as ``quasi-ideal'', the kinetic energy of the
condensate is large compared to the strength of the external potential, and
interactions tend to be negligible. We find, in agreement with previous
theoretical studies \cite{Leb01,Pav02}, a strong suppression of dissipation as
the velocity increases.
The condensate density is again only locally distorted in the vicinity of the
defect (cf Fig. \ref{F2}(d)). 
This local distortion is very well described
by applying
the combination of perturbative and local density approach already used in the
SF regime. The density is of the same form as in Eq. (\ref{eq}) with here
\cite{Leb01}
\begin{equation}\label{80}
\delta n(x,t)=\frac{4\,m}{\hbar^2\,\kappa} \left[\frac{U(x)}{\kappa}
+\Theta(-x\dot{X}_t) 
\, \Im \left\{e^{i\kappa x} \hat{U}(\kappa)\right\}\right] \, ,
\end{equation}
where $\hat{U}$ is the Fourier transform of $U$.  This form indeed
corresponds to almost perfect transmission, with small reflexion on the
defect (see the inset of Fig. \ref{F2}(d)), the amount of which decreases
at large velocity ($\kappa\to\infty$).

The frontier between the dissipative and the quasi-ideal region
can be estimated by studying the related problem of a homogeneous condensate
flowing through a barrier potential \cite{foot}.
In this simplified configuration it is possible to determine analytically the
velocity at which the system undergoes a transition from a local perturbation
to an irregular fluctuating density profile \cite{Leb01}.  We find that this
estimate fits very well the numerically determined supersonic frontier (see
Fig.~\ref{F1}). This stresses the qualitative similarities between
Fig.~\ref{F1} and the phase diagram obtained for a defect moving through a
homogeneous fluid \cite{Leb01,Pav02}.  Interestingly, the existence of the
three regions (SF, nonlinear dissipative, and quasi-ideal weakly-damped) was quite
recently observed experimentally for a localized defect in
Ref. \cite{Engels07}.

\begin{figure}
  \includegraphics[width=0.8\linewidth]{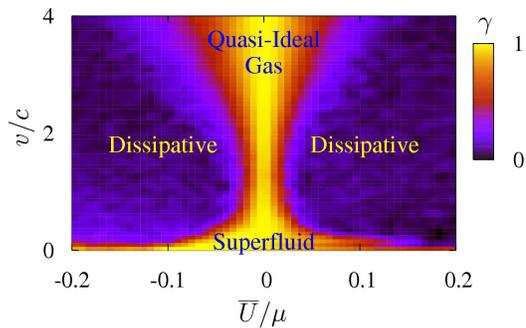}
  \caption{(color online) Fluidity factor $\gamma$ of dipole oscillations in
    presence of a speckle potential.}
  \label{phase_diag_d}
\end{figure}

{\sl Disordered potential.} Keeping the same parameters as in Fig. \ref{F1}, we
now replace the single localized impurity by a disordered potential and
compute, as before, the fluidity factor as a function of velocity and of the
intensity of the (now random) potential.  Fig.~\ref{phase_diag_d} corresponds
to the case where $U(x)$ is an optical speckle potential of mean value
$\overline U$ with a correlation length $l_c$ such that $L/l_c=30$, typical
of experimental configurations \cite{bemerkung0}. The picture is generic 
and does not depend of the details of the disorder. The main result that
emerges from the comparison of Fig.~\ref{F1} and Fig.~\ref{phase_diag_d} is
that, in the weak--disorder limit that we explore (${\overline U}/\mu \ll 1$), the global 
properties of the damping phase diagram are qualitatively similar in both cases. 
The same three phases observed for the localized defect are again present. 
However, their relative importance is quite different. One observes a considerable 
shrinking of the SF and quasi-ideal weakly-damped regions, compared to the nonlinear 
dissipative one.

In the presence of disorder, important experimental efforts have been
undertaken for understanding the possible connection of damping of dipole
oscillations with Anderson localization of matter waves. Our analysis shows that
damping occurs in the dissipative phase. However, the dissipative mechanism
observed in this phase is --as in the case of a localized defect--
connected to the loss of collectivity due to the emission of solitons and linear 
excitations, and not to a localization phenomenon. From what is known in the case 
of an a ho\-mo\-ge\-neous flow, genuine Anderson localization might only occur
in the deep supersonic regime $v/c \gg 1$ \cite{Paul07}. Without
going into a detailed analysis of the experimental results
\cite{Lye05,Chen07}, let us mention that the displacement of the harmonic
potential (around 700 $\mu$m) used at Rice University corresponds to $v/c
\approx 2.8$, while the lowest speckle height considered is
$\overline{U}/\mu \approx 0.04$, which locates the system in the dissipative
phase (under the experimental conditions, we find that at $v/c \approx 2.8$
the dissipative phase border is at $\overline{U}/\mu \approx 0.008$). Similarly, from the
data published in Ref.~\cite{Lye05} our analysis shows that the experiments at
Florence were also performed in the supersonic dissipative region. These simple remarks
explain the experimentally observed damping, and locate the experimental
configurations in a regime where the effects of Anderson localization are at
best indirect.

\begin{figure}
  \includegraphics[width=0.80\linewidth]{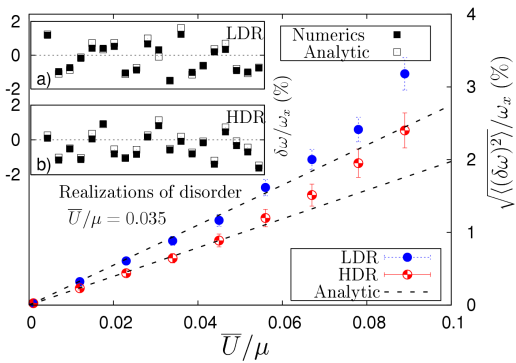}
  \caption{(color online) Relative standard deviation $ \omega_x^{-1}
    \langle\delta\omega^2\rangle^{1/2}$ as function of average speckle
    intensity. The analytical results correspond to Eq.  (\ref{rms_shift}).
    The insets compare numerical and analytical frequency shifts for different
    realization of disorder for $\overline{U}/\mu=0.035$ in the HDR and LDR.}
  \label{fluctu}
\end{figure}

In the subsonic SF regime it is possible to compute the frequency shift due to
the disordered potential similarly as in the one-peak case. The approach is
found to be equivalent to the sum rule approach developed in
\cite{Lye05,Modugno06}. 
One can also derive a simple relation for the variance of the frequency shift:
\begin{equation}\label{rms_shift}
  \langle (\delta \omega)^2\rangle=
\iint_{\mathbb R^2} \frac{dx  dx'}{4 m^2 \omega_x^2} 
\,\frac{d^2 n_0(x)}{dx^2}\,\frac{d^2 n_0(x')}{dx'^2} \langle U(x)U(x')\rangle
\; ,
\end{equation}
where $\langle \dots \rangle$ denotes ensemble average. For small
$\overline{U}/\mu$ (when the lowest order approximation holds)
Eq.~(\ref{rms_shift}) is found to be in very good agreement with numerical
integration of the GP equation, as seen in Fig. \ref{fluctu}.

To conclude, we have presented a comprehensive picture of the damping
properties of dipole oscillations of BEC in the presence of a scattering
potential. Strong analogies are stressed between different types of
potentials. Three different phases are shown to exist: superfluid ($v/c <
1$), nonlinear dissipative ($v/c \sim 1$) and quasi-ideal ($v/c
> 1$). The mechanism that breaks SF and leads to damped oscillations is
shown to correspond to a generic onset of dissipation which, in the presence
of disorder, is unrelated to localization properties.  As the strength of the
disorder potential increases the nonlinear dissipative phase occupies most of
the phase diagram. Our findings allow to give a simple
interpretation of experimental results.

This work was supported by grants ANR--05--Nano--008--02 and
ANR--NT05--2--42103, by the IFRAF Institute and by the Alexander von Humboldt
Foundation. We are grateful to L. Sanchez-Palencia for discussions.


\begin{thebibliography}{99}
\bibitem{foot} In the deep supersonic regime the moving
condensate can be considered in the vicinity of the obstacle
as a stationary homogeneous flow because
the defect produces only a local perturbation, see Eq. (\ref{80}).
\bibitem{bemerkung0} The intensity of the potential is usually characterized 
by $U_D = 2 |\overline U|$ \cite{Lye05,Chen07}.
\end{thebibliography}
\end{document}